\documentclass[12pt,preprint]{aastex} 
 
\tighten 
 
\slugcomment{Accepted for publication in {\it the Astrophysical Journal Letters} 2002 July 24 } 
\def\lsim{\lower.5ex\hbox{$\; \buildrel < \over \sim \;$}} 
\def\gsim{\lower.5ex\hbox{$\; \buildrel > \over \sim \;$}} 
\def\lax {\ifmmode{_<\atop^{\sim}}\else{${_<\atop^{\sim}}$}\fi} 
\def\gax {\ifmmode{_>\atop^{\sim}}\else{${_>\atop^{\sim}}$}\fi} 
 
\def\gtorder{\mathrel{\raise.3ex\hbox{$>$}\mkern-14mu 
\lower0.6ex\hbox{$\sim$}}} 
\def\ltorder{\mathrel{\raise.3ex\hbox{$<$}\mkern-14mu 
\lower0.6ex\hbox{$\sim$}}}

\def\pmb#1{\setbox0=\hbox{#1}%
\kern-0.015em\copy0\kern-\wd0 
\kern0.03em\copy0\kern-\wd0 
\kern-0.015em\raise0.0433em\box0 } 
 
\begin{document} 
 
\title{Why is it Difficult to Detect  a Millisecond Pulsar in Neutron Star 
X-ray Binaries?} 

\author{Lev Titarchuk \altaffilmark{1,3}, Wei Cui\altaffilmark{2}
and Kent Wood \altaffilmark{3}}

\altaffiltext{1}{George Mason University/Center for Earth Observing and
Space Research, Fairfax VA 22030-4444}
\altaffiltext{2}{Department of Physics, Purdue University, West Lafayette,
IN 47907; cui@physics.purdue.edu}
\altaffiltext{3}{US Naval Research Laboratory,  Space Science
Division, 4555 Overlook Avenue, SW, Washington,
DC 20375-5352; lev@xip.nrl.navy.mil; kwood@ssd5.nrl.navy.mil}

\vskip 0.5 truecm 
 

\begin{abstract} 
We explain why it is possible to detect directly X-ray emission from near the 
surface of the neutron star (NS) in SAX J1808.4-3658 but not in most other
low-mass X-ray binaries (LMXBs), with the exception that 
 emission from the surface can  be  seen 
during bursts events. We show that the X-ray emission from 
SAX J1808.4-3658 mostly originates in the Comptonization process 
in a relatively optical thin hot region (with an optical depth $\tau_0$ 
around 4 and temperature is around 20 keV). Such a transparent region 
does not prevent us from detecting coherent X-ray pulsation due to hot 
spots on the NS surface. 
We give a precise model for the loss of modulation,
such suppression of the QPO amplitude due to scattering can explain
the disappearance of kHz QPOs with increasing QPO frequency. 
We also formulate general conditions under which the millisecond X-ray 
pulsation can be detected in LMXBs. 
We demonstrate that the observed soft phase lag of the pulsed emission
is a result of the downscattering of the hard X-ray photons in the 
relatively cold material near the NS surface.
In the framework of this downscattering model we propose 
a method to determine the atmosphere density in that  region from soft-lag
measurements.
\end{abstract} 
 
 
 
\section{Introduction} 
Low-mass X-ray binaries (LMXB) presumably contain 
a  weakly magnetized neutron star (NS).
Recently Titarchuk, Bradshaw \& Wood (2001), hereafter TBW 
suggested a new method of estimating B-field strength  using
the magnetoacoustic oscillation model and  found that 
the B-field strength for a number 
of NSs in  LMXBs is  $\sim 10^8$ gauss.  
Near-coherent millisecond X-ray pulsations have been 
observed in 4U 1728-34
during thermonuclear (type I) X-ray bursts (e.g. Strohmayer et al. 1996).
They are interpreted as X-ray intensity  modulated at period close to the spin period of the neutron star. 
Thus these B-field estimates in addition to the detection of the milisecond pulsation 
give strong arguments for LMXB neutron stars to be   progenitors of millisecond radio pulsars (MLP)
(see review by Bhattacharya \& van den Heuvel 1991). 
 But there is still a question why these coherent pulsations are not found in persistent emission
 despite careful searches (Wood et al. 1991; Vaughan et al. 1994).
The lack of coherent pulsations has been explained as modulation loss from 
 gravitational lensing (Wood, Ftaclas \& Kearney 1988; Meszaros, Riffert \&
Berthiaume 1988) or from to scattering (e.g. Bainerd \& Lamb 1987; Kylafis \& Klimmis 1987 ).
The third explanation  for  the lack  of pulsations is presented by 
Cumming, Zweibel \& Bildsten (2001) who argue that the surface field is weak because of magnetic
screening  \footnote{
The recent discoveries of MLP in XTE J1751-305 (Markwardt et al. 2002) and in 
XTE J0929-314 (Galloway et al. 2002 ) with extremely low mass transfer rates support the suggestion 
of the absence  of the magnetic screening  for the  low mass accretion rates.}.
In this {\it Letter} we put forth arguments for smearing out of the pulsar signal due to 
electron scattering in the optically thick environment typical of  most  observed 
LMXBs.
In  \S 2 we study the scattering effect and its relation with
the observed timing and spectral characteristics  for various QPO sources.
In \S 3  we investigate the scattering effects in QPO sources.
In \S 4 we present results of the RXTE data analysis of spectral  properties of X-ray radiation in   
SAX 1808.4-3656. In \S 5 we analyze the downscattering model and its application to the observed
soft lag phenomenon detected in the coherent pulse signal from SAX 1808.4-3656.
Conclusions follow in \S 6. 

\section{Scattering Effect on  QPO Pulse Profile and Pulsar Profile}  

The transfer problem to be solved is that of a sinusoid, representing either (a)
an intrinsic pulsar light curve or (b) a high frequency QPO, but in each case
seen through a surrounding medium that reduces the pulse modulation 
by scattering. The signal whose modulation is affected lies in the range
of a few hundred Hz to perhaps 1 kHz. We thus must analyze smearing out of a signal
at very high frequencies and investigate the dependence upon both the intrinsic
frequency and the optical depth of the scattering environment. The geometry used
derives from earlier works (see e.g. Sunyaev \& Titarchuk 1980, hereafter ST80;
Titarchuk, Lapidus \& Muslimov 1998, hereafter TLM98)
modeling energy spectra and power density
spectra (PDS) of LMXB sources. 
{\it For the high frequency QPO,  [case (b)]
the TLM98 model identifies the site where the QPO is generated}, which will be at radius substantially
larger than that of the star. Thus the effective optical depth of interest in case (b) will not be
the full optical depth to the star surface but rather the reduced optical depth appropriate to that
site location.

One more consideration completes the problem specification.
It is well established from observation that the kHz frequency QPOs in LMXBs are not fixed in frequency
but increase or decrease in manner correlated with changes in luminosity.
which in turn is believed
to vary directly as mass accretion rate, hence as source brightens the OPO site moves both inward in
radius and upward in frequency. 
Consequently, 
the total change in QPO power as the source varies
will be a combination of two effects, namely (i) the intrinsic change in QPO power that results
simply from physics of shifting the resonance frequency that gives the QPO and (ii) the further
reduction in power that results from its now being {\it observed at higher frequency and through
 greater optical depth}, both of which enhance the effects of scattering.  
To model the entire process we therefore require a model for the intrinsic change in power.
This is provided in the cited transition layer (TL) model (TLM98)
by approximating the QPO process itself as
a damped harmonic oscillator under the influence of a driving force. Then, the interaction leads to
resonance oscillations for which the Q-value $(\nu_{QPO}/\delta\nu_{QPO})$ can be high.
The resonance amplitude varies as the product of the amplitude of the driving force $A_{dr}(\nu)$ and
$1/\nu_{QPO}$ (see Landau \& Lifshitz 1965). The default intrinsic model, in the absence
of scattering, is thus that the QPO amplitude  falls off as  $A(\nu)=A_{dr}(\nu)/\nu_{QPO}$, but scattering
contributes additional reduction. In a source such as GX 340+0, where a {\it pair} of high frequency
QPOs is seen, it is possible to test out these ideas in detail, using the model derived below and
applying it to both QPOs, keeping the optical depth $\tau_0$ the same for both kHz QPOs because they
are co-located.
Detection of the NS spin period is the crucial further issue at stake here.
The typical problem is the following: since the NS spin gives a {\it coherent} periodic signal at
nearly constant frequency 
it should be at relative
 {\it advantage}  in detectability in comparison with QPOs, yet is not see while the QPOs {\it are
 seen}. 
 The key, obviously, is that the QPO sees the
 reduced optical depth while the modulation at the spin period experiences the unreduced optical depth.
 The occasional exceptions where the stellar spin frequency does prove detectable (such as MLPs,
  plus the transient detections in X-ray bursters) should be explicable in terms of
 circumstances that reduce scattering effects in those particular cases below values that generally
 obtain; moreover those exceptional reductions should be confirmed by spectra indicative of the lower
 optical depth and  other supporting context observations.

For a quantitative model of this problem scattering  is handled as a diffusion problem in a Green's
function [$G(t)$] treatment.
The input  pulse is $x(t)=A\sin({\omega
t}+\varphi_0)$ where $\omega=2\pi \nu$ is the NS spin rotational  frequency $\omega$ or the QPO 
centroid frequency depending on whether it case (a) or (b) and 
$\varphi_0$ is an initial phase.
The resulting pulse affected by the scattering is $z(t)$ and is given by a convolution of $x(t)$ and 
$G(t)$:
\begin{equation}
 z(t)=\int_0^t x(t^{\prime})G(t-t^{\prime})dt^{\prime}.
\end{equation}   
For times $t$ much greater than the scattering time in the Compton cloud 
$R\tau/c$ the initial pulse emission $x(t)=A(\omega)\exp{[(\omega t+\varphi_0)i]}$ is
calculated as follows
\begin{equation}
z(t)=A(\omega)\exp{[i(\omega t+\varphi_0)]}I(\omega),~~~~
{\rm and}  ~~~~
I(\omega)=\int_0^{\infty}e^{-i\omega u}G(u)du,~~~
\end{equation}
ST80
analyzed the response of the scattering medium (the Green's function)
for various source distributions. 
They show that
for a QPO source embedded in the center of the cloud ($\tau=0$)
the response of the scattering medium (the Green's function), $G(t)$ 
can be approximated by 
\begin{equation}
 G(t)\propto(C_1+C_2t^{-5/2})\exp[
 -3R\tau_0/4ct
 -\pi^2c\tau_0t/3R(\tau_0+2/3)^2],
 \end{equation}
 where $\tau_0$, $R$ are  optical depth and radius of the cloud respectively,
 $c$ is the light speed. Coefficients $C_1$ and $C_2$ can be estimated as
 follows (see also Wood et al. 2001)
 $C_1=5\pi^2/3(\tau_0+2/3)^2$, 
and 
 $C_2=\pi(3/\pi)^{3/2}\tau_0^{1/2}
 (R/c)^{5/2}/4$.
Using formula (3) the integral  $I(\omega)$  is analytically calculated  
 by the steepest descent method 
 (see also  Prudnikov, Bruchkov \& Marichev 1981, formula 2.3.16).
 We can show  that an amplitude $B(\omega)$ of $z(t)$
 can be obtained as
 \begin{equation}
 B(\omega)= A(\omega){\cal F}({\cal X})
 \exp{ (-2\cal X)}
 \end{equation}  
 where ${\cal X}(\omega,\tau_0)=\sqrt{[\pi\tau_0/2(\tau_0+2/3)]^4+(3R\tau_0 
 \omega/4c)^2}$ and ${\cal F}({\cal X})=d_1{\cal X}^{1/2}+
 d_2{\cal X}^{-3/4}+d_0$ is a weak function of
 ${\cal X}$
 which is a constant  and the product of ${\cal F}({\cal X})\exp{ (-2\cal X)}$ is almost unity
 for $\omega$ much less than the inverse of crossing time 
 $t_0^{-1}=(r\tau_0/c)^{-1}$.  
 The amplitude $B(\omega)$  decays exponentially when $\omega$ increases.
 
\section{Consequencies  of scattering effects for QPO observations:
Evidence of the resonance excitation in the QPO sources}  
 
As it is seen from  formula (4) the  amplitude of the NS 
pulsations with $\omega\sim 2\pi \times 400$ Hz  
decreases very rapidly with $\tau_0$ for 
\begin{equation}
\tau_0~~\gax ~~4[(c/R)/5\times 10^3~{\rm s}^{-1}]/[\omega/(2\pi\times 400~{\rm Hz})].
\end{equation}
A radius $R$ of 60 km was chosen as a typical radius of
the Compton cloud derived from the observations in the frameworks of the TL
model (see TBW). 
In Figure 2 we present the rms amplitude of the high frequency QPOs as a function of
kHz frequencies for the Z-source GX 340+0 using the results of  Jonker et al. (2000). 
We compare them with the theoretical model presented by formula (4).
The amplitude of the lower kHz QPOs is perfectly fitted by $1/\omega$ dependence (see 
the dashed line on the left
side of Fig. 2). $\chi^2_{red}=0.53$ for this fit,
but the deviation of $B(\omega)$ due to  scattering (the  solid line on the right side of Fig. 2) from the  
$1/\omega$ law (the dashed line on the right side of Fig. 2) is  clearly seen for the higher kHz QPOs.
The best fit to the data ($\chi^2_{red}=1.1$) by formula (5) with $A(\omega)\propto 1/\omega$ is obtained
for $\tau_0=2.7$ with $R/c=10^{-4}$ s. 
For $\tau_0=2.7$ the rms, $B(\omega)$ calculated  for low kHz frequencies is almost identical
to $1/\omega$ dependence ($\chi^2_{red}=0.85$).
Thus we found that  the driving force amplitude $A_{dr}$ is practically independent of frequency for
kHz QPOs. 
One can argue that the driving force power for kHz QPOs  can be determined by the observed PDS at kHz  
frequencies. The PDS power law index $\beta$ typically varies  around 1
(see details in Wijnands \& van der Klis 1998). 
The resulting resonance amplitude would be $\omega^{-(1+\beta/2)}$ with an assumption that this power 
continuum is a driving force reservoir for the resonance. As an example we fit 
this $\omega^{-(1+\beta/2)}$ model with $\beta=1$  to the data for the {\it lower} kHz QPOs 
only, where the scattering effects can be neglected (see inequality 5). 
We found that   for this model $\chi^2_{red}=1.37$. 
The rms-frequency dependence  for the lower kHz QPOs 
is not affected by scattering; it affects mainly the higher kHz QPOs (see Eq. 5).
The scattering effects make the rms-frequency dependence much steeper for the higher kHz QPOs which does not allow us
to extract the true driving force amplitude from the data (clearly, it is not feasible
to dispense with scattering and fit both QPOs solely with type of driving force model)
\footnote{Fits of the rms amplitude vs frequency similar to that presented in Fig. 2 for GX 340+0, have been produced
for GX 5-1, Cyg X-2, GX 17-2 and Sco X-1.
Thus one can conclude that in all these kHz QPO observations  
the rms resonance amplitude $1/\omega$ vs frequency    is affected by the photon scattering.}.
It is well established in the QPO observations that  there is a correlation between the QPO frequency
and the mass accretion (or the count rate). The high QPO frequency increases with the count rate
reaching a certain level and then it disappears as the low frequencies are still detected 
(Zhang et al. 1998; Mendez et al. 1999).
  This effect can be explained by the exponential decay
of the QPO amplitude with increase of $\tau_0\omega$ which is a result of the simultanenous rise of
$\tau_0$ and $\omega$ with the mass accretion rate. (see Eq. 4).   
The high frequency QPOs  presumably originate  in the outskirts 
of the Compton cloud and 
the optical depth of the QPO source location is relatively small (1-3) 
even in systems  with high accretion rate. 
In contrast,  the source of the NS pulsations 
is deeply embedded in the Compton cloud, whose  
For high mass accretion  it is hardly to expect 
the detection of the pulsations from the NS surface. The Compton cloud 
optical depth can be very
large. The NS pulsations would be totally smeared out by
the scattering  if the pulsation frequency is about 300-400 Hz and higher.  
In the next section we show that in SAX J1808.4-3658 the optical depth of spherical Compton cloud is
relatively low allowing detections.

\section{X-ray spectra of SAX J1808.4-3658 and of Other LMXBs}  
 
We revisit some  RXTE spectral data for SAX J1808.4-3658 (Gilfanov 
et al. 1998; Heindl \& Smith 1998). The spectrum (which was
constructed using data from 30411-01-06-00) is well
represented by the Comptonization model (CompTT in XSPEC, with spherical
geometry; Titarchuk 1994)
 along with some 
contribution from a blackbody
component and a Fe $K_{\alpha}-$ line.
The best-fit parameter of the model are $\tau_0=4.5^{+0.55}_{-0.97}$, electron temperature 
$kT_e=19.5^{+7.8}_{-3.2}$  keV,  $K_{\alpha}$ energy is at $6.5\pm 0.1$ keV. 
The temperature of the seed photons for the Comptonization is within $
kT_0=0.1^{+0.4}_{-0.1}$ keV.
The temperature of the blackbody (BB) radiation (presumably from the
neutron star surface) is
$kT_{bb}=0.69\pm0.02$ keV. The quality of the fit is high with 
$\chi^2=294/322=0.91$
\footnote{Gierli\`nski, Done \& Barret (2002), hereafter GDB, also analyzed
the X-ray
spectrum of SAX J1808.4-3658 and they showed the spectrum can be fitted by
a three-component model (BB component,  Comptonization component and the 
Compton reflection).
In general, spectral modeling is not unique  
and it is not by chance that our spectral model and  global picture are  different from that of the GDB.
Because at $kT_{bb}$ around 0.7 keV the iron is highly ionized we   conclude that the observed $K_\alpha$ 
(see \S 4 and GDB) is formed as a result of the disk reflection rather than  as  the NS  reflection.
 }.
It is worth noting the relatively small optical depth $\tau=4.5$.
This can be the main reason why  the NS pulsation is detected from this source
(see inequality 5).
If  this $\tau_0$ is compared 
with that for other LMXBs sources (showing QPO features) one can find that
optical depths for a number of sources are   higher than that for  
SAX J1808.4-3658. They are 10-11 for Cyg X-2 (Kuznetsov 2002), 
Sco X-1 and  GX 340+0,  $\tau_0$ is around 6 for  4U 1728-34 (TBW). 
 On average the seed photon temperature value 
 $T_0\ll T_{bb}$. Presumably the photon supply for Comptonization
mostly comes from the relatively cold disk and the blackbody emission comes 
from the NS surface. 
The Comptonization  spectra of the QPO sources where the NS pulsations are
not observed is  indirect evidence for this. 
{
van der Klis (2000) argues
that there is a striking similarity between the power spectra of black holes, atoll sources, 
Z sources and the MLP 
SAX J1808.4-3658. This would exclude any spectral formation models requiring a
material surface and would essentially imply the phenomena are generated in the
accretion disk around any low-magnetic field compact object.
But in the case of  SAX J1808.4-3658 we definitely 
see  coherent pulsations of the hard radiation
which  originates in the NS surface [Cui, Morgan \& Titarchuk (1998), hereafter CMT98] and also
we should see the downscattering effect of the hard Comptonized photons
when they  hit the relatively cold material of the NS star (see Fig.1). 

\section{Downscattering effect}
For SAX J1808.4-3658, the observed characteristics of the pulsed X-ray emission
(pulse profiles and phase lags) and the overall energy spectrum provide
useful  insight into X-ray production processes and the emission environment.
 Important information regarding the density of the NS atmosphere   
can be extracted from the time lags between different energy 
bands. The soft lags have been detected for SAX J1808.4-3658
by  CMT98\footnote{Galloway et al. 2002 also found the hard X-ray pulses arrived up to 
770 $\mu$s earlier the soft X-ray pulses in  XTE J0929-314.}.
The soft phase lag can be  due to reflection  (Compton downscattering)
of the pulsed hard radiation from the NS atmosphere.
In fact, this effect is  unavoidable when hot plasma streams converges
towards the NS magnetic poles. The hard radiation at the pole illuminates 
adjoining regions where a significant fraction of the hard radiation is 
reflected.
For simplicity we assume the input hard photons
are monochromatic with energy $E_h$. The photons that emerge from 
the NS atmosphere with lower energy $E_l$ arrive at distant observer later
than those with higher energy $E_h$. {\it The delay in arrival time, $\delta t$,
is given by $u l/c$, where $u$ is the number of
scatterings undergone by  the seed hard photon with energy $E_h$} before emerging
with energy $E_l$ and $l=1/n_e\sigma_T$ is the Thomson mean free path for the
photon.
The average fractional energy loss for the photon of energy $E$ 
after each scattering is nearly independent of $T_{bb}$ and it is given by
$\delta E/E\approx -(E-4kT_{bb})/m_ec^2\approx -E/m_ec^2$ (because $4kT_{bb}\ll E$), 
where $m_ec^2$ is the electron rest mass.
Integrating over multiple scatterings, we have $u=m_ec^2(1/E_l-1/E_h)$
(ST80).
The measured soft lag  scales as $E^{-1}$ as shown in Figure 3, and
levels off above 10 keV. 
In Figure 3 we also show the best fit to the data using the downscattering
model 
\begin{equation}
E^{-1}=E_h^{-1}+D\cdot\delta t.
\end{equation} 
We fit the theoretical  $E^{-1}$ as a function of $\delta t$ to the measured
quantities.
The best-fit parameters are $E_h=20$ keV and $D=1.3\times10^{-3}$ $\mu$s$^{-1}$ 
keV$^{-1}$ for which  $\chi^2_{red}=11.3/9=1.2$.
Because $D=n_ec\sigma_T/m_ec^2$ we find that {\it the NS density 
near  the surface is} $n_e=Dm_ec^2/c\sigma_T=
3.3\times10^{19}$ cm$^{-3}$.
The measured soft-lags in the frameworks of the downscattering models allow us to estimate 
the optical depth  $\tau$ up to what  the hard photon can penetrate illuminating
 the NS atmosphere. 
It takes more than 70 scatterings for 20 keV photon to reach $~5$ keV energy band. As it is
known from the diffusion theory  (see e.g. ST80) a  photon in the course of the random
diffusion propagation to optical depth $\tau$ undergoes on average $N_{sc}\sim\tau^2$ 
scatterings. Thus using  $N_{sc}=70$ 
(the scattering number of the input photons which finally emerge) one can deduce the penetration optical depth 
$\tau_{pr}=\sqrt {N_{sc}/2}=6$.  In other words  the thickness of 
the NS reflection slab is approximately $(m_p/\sigma_T)\tau_{pr}=15$ grams cm$^{-2}$.    
Photoelectric absorption is suppressed at $kT_{bb}=0.7$ keV 
and thus one can estimate $\tau_{pr}$ neglecting the photoelectic extinction. 
The deduced $\tau_{pr}$ leads to estimate of the
reflection albedo $A=1-(3\tau_{pr}/4+1)^{-1}=0.82$ (see e.g. Sobolev 1975). 
It means that more than 80\% of the input hard photons are reflected by the NS atmosphere. 

\section{Conclusions} 
Our analysis of the timing and spectral properties of milisecond pulsar 
SAX J1808.4-3658 and comparison of them with those for LMXB QPO sources
leads us to these conclusions: 
 (1)  the detection of the NS pulsations from SAX J1808.4-3658 
 was possible because of the relatively transparent Compton cloud covering
 in this source ($\tau_0\sim 4$). For the majority of the analyzed LMXBs the Compton cloud optical depth is at
 least twice as high as that in SAX J1808.4-3658.
 The high frequency pulsations with frequencies 
 $\gax 300$ Hz are strongly wiped out by scattering in the clouds with 
 $\tau_0>4$.  SAX J1808.4-3658 is a limiting case for this detection. 
 There is a possibility of finding the NS pulsations in the
 sources that are much less luminous (than these  bright QPO LMXBs), because  in them 
 the Compton cloud is more  transparent ($\tau_0\lax 4$) for the NS pulsed
 radiation. (2) The kHz QPOs  rms vs frequency  follows the resonance law
$1/\omega$ are weakened  by  scattering in the Compton cloud. 
 (3) The soft-lag measurements along with the implementation of
the downscattering model  can provide a tool  
for  density determination near  the NS 
surface.  

We appreciate the fruitful discussions with   Paul Ray, Sergey Kuznetsov and
 particularly, with the referee.


\clearpage

\begin{figure} 
\epsscale{0.7} 
\plotone{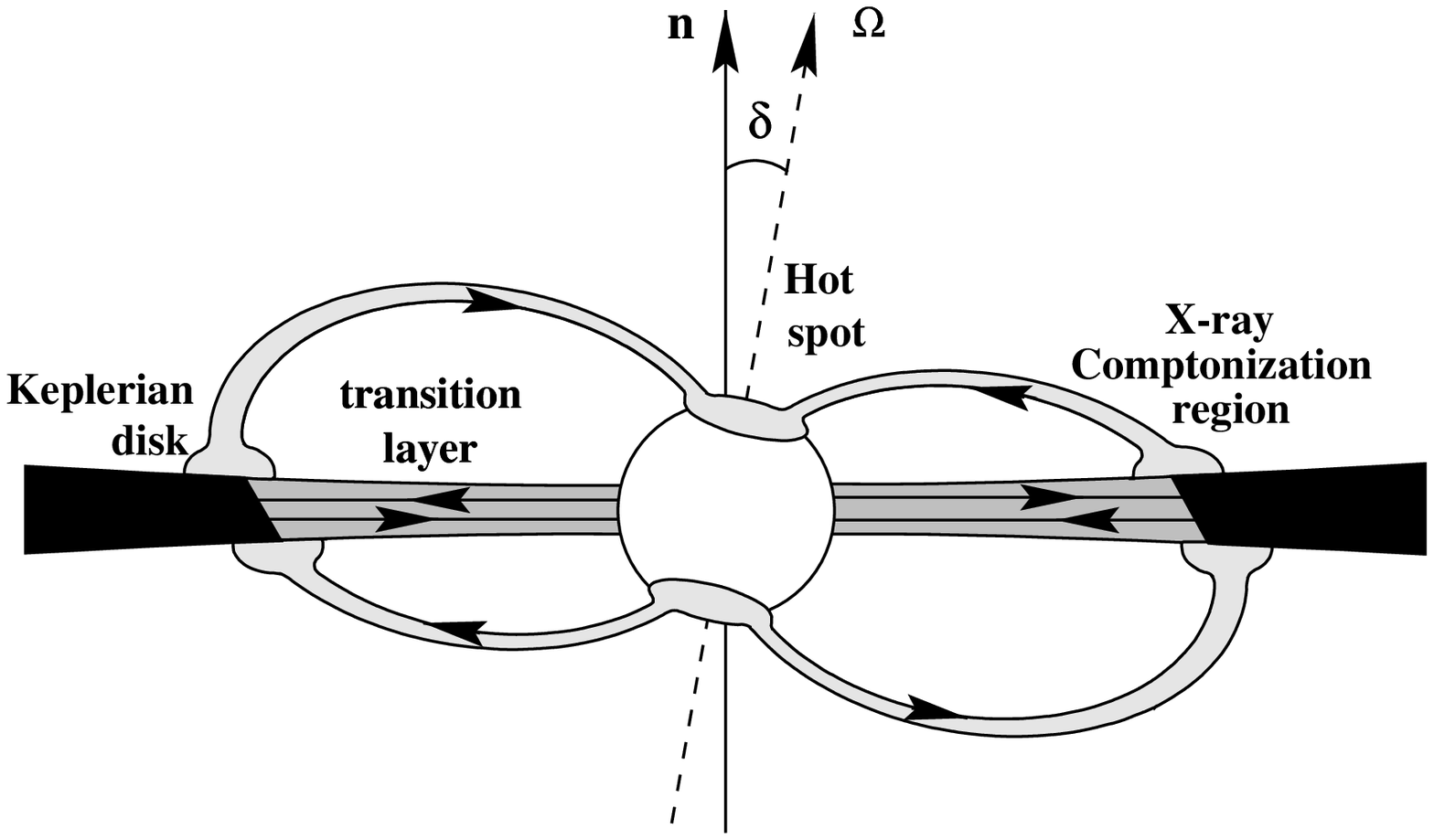} 
\caption{Schematic picture of the model geometry SAX J1808.4-3658.
The hot flow forming at the outer boundary of the transition layer streams
towards neutron star along magnetic field lines. The reflection of the hard
radiation of the flow from the neutron star surface results in soft time lags
of the pulsed emission.} 
\label{fig1} 
\end{figure} 

\begin{figure} 
\epsscale{0.7} 
\plotone{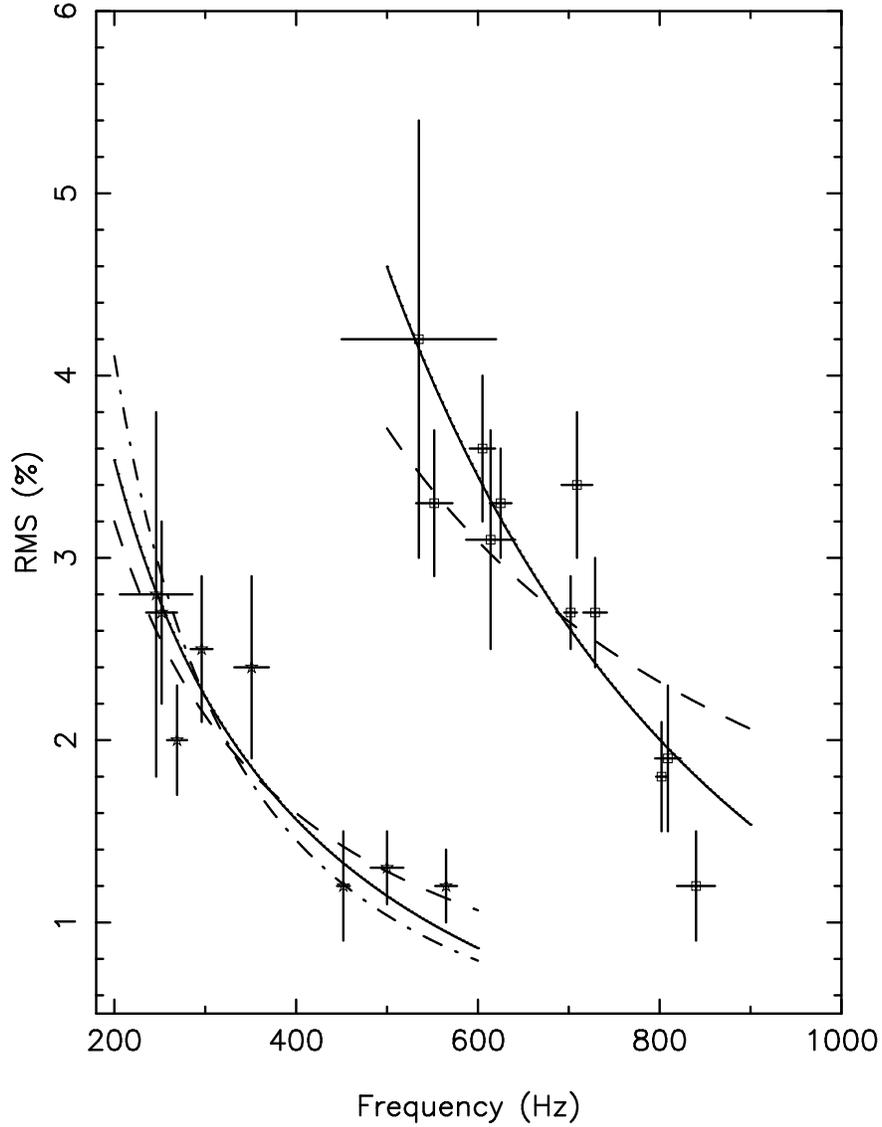} 
\caption{ Examples of RMS amplitude vs QPO frequency for lower and higher kHz peaks
for GX 340+0 (Jonker et al. 2000). Two solid lines (for lower and higher peaks respectively) 
are theoretical curves for which  scattering effects  are taken into account. Dashed lines are for $1/\nu$ law 
and dash-dotted  line is  for  $\nu^{-3/2}$  law.
}
\label{fig2} 
\end{figure} 

 
 
\begin{figure} 
\epsscale{0.7} 
\plotone{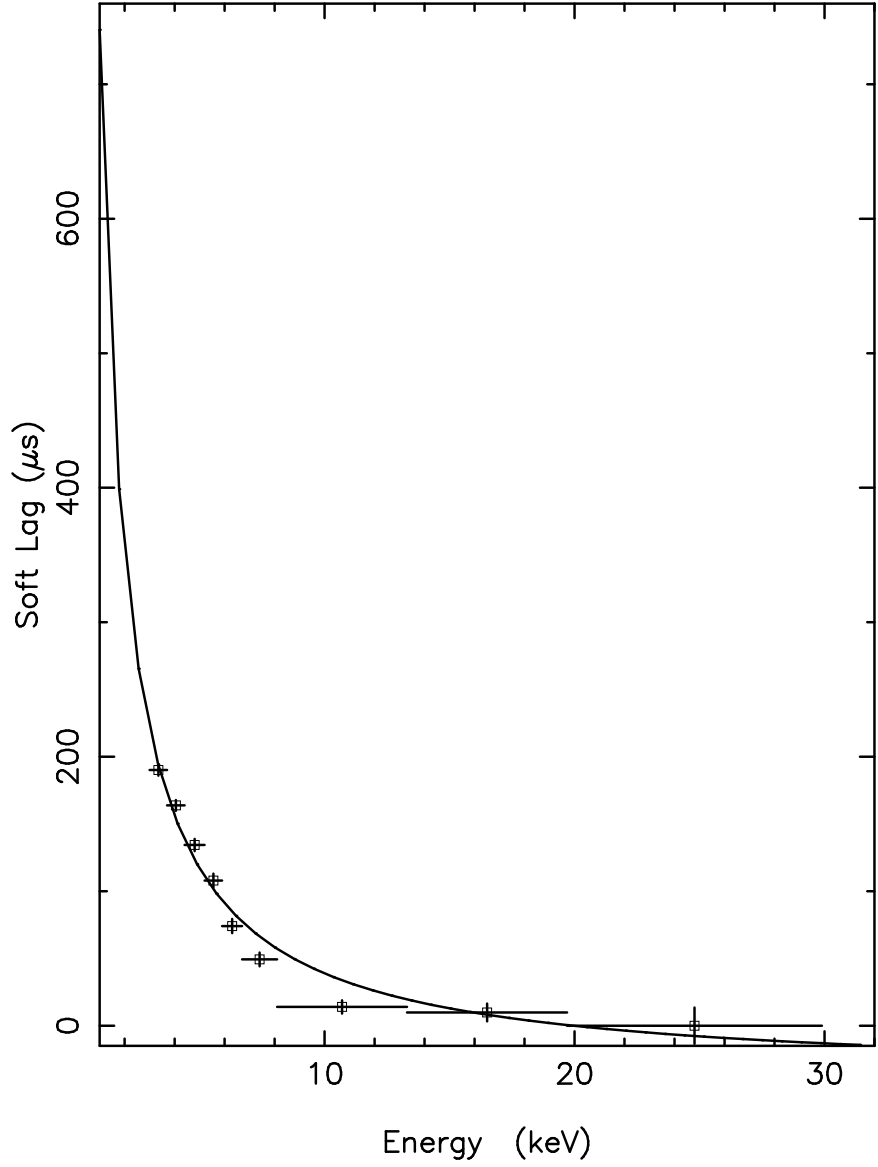} 
\caption{Measured soft X-ray lags with respect to 20-30 keV band along with 
the best-fit downscattering model.} 
\label{fig3} 
\end{figure}

\clearpage 
 
\end{document}